\documentclass{llncs}

\usepackage{amsmath,amsfonts}

\usepackage{amsthm}
\usepackage{thm-restate}
\theoremstyle{definition}

\usepackage{todonotes}
\usepackage{algorithm}
\usepackage[linesnumbered,ruled,algo2e,noend]{algorithm2e}
\SetAlgoCaptionSeparator{}

\usepackage{caption}
\usepackage{graphicx}
\usepackage{textcomp}
\usepackage{xcolor}
\usepackage{multirow}
\usepackage{booktabs}
\usepackage{enumitem}
\usepackage[scaled=0.85]{beramono}
\usepackage[T1]{fontenc}
\usepackage[utf8]{inputenc}
\usepackage{bm}
\usepackage{wrapfig}
\usepackage{tikz}
\usetikzlibrary{arrows,automata,positioning,calc,arrows,arrows.meta}
\usepackage{flushend}
\usepackage{listings}
\usepackage{multirow}
\usepackage{hyperref}

\usepackage{soul}
\setstcolor{red}%
\setul{}{2pt}

\newcommand{\threatmodel}{M}

\newcommand{\elements}{E}
\newcommand{\assets}{A}
\newcommand{\connectors}{C}
\newcommand{\boundaries}{B}
\newcommand{\items}{I}
\newcommand{\elementtypes}{\ensuremath{\mathbb{E}}\xspace}
\newcommand{\assettypes}{\ensuremath{\mathbb{A}}\xspace}
\newcommand{\connectortypes}{\ensuremath{\mathbb{C}}\xspace}

\newcommand{\boundarytypes}{\ensuremath{\mathbb{B}}\xspace}
\newcommand{\itemtypes}{\ensuremath{\mathbb{I}}\xspace}

\newcommand{\source}{\footnotesize\texttt{src}}
\newcommand{\target}{\footnotesize\texttt{tgt}}

\newcommand{\attribute}{\mathfrak{a}}
\newcommand{\attributedomain}{\mathfrak{D}}
\newcommand{\itemattribute}{\mathbf{a}}
\newcommand{\domainofattribute}{\mathfrak{D}}
\newcommand{\attributevalueassignment}{v}
\newcommand{\attributeweight}{w}

\newcommand{\attributeuniverse}{\mathfrak{A}}
\newcommand{\attributevalueuniverse}{\mathfrak{V}}

\newcommand{\crosses}{{\footnotesize\texttt{crosses}}}

\newcommand{\contains}{{\footnotesize\texttt{contained}}}
\newcommand{\type}{\tau}

\newcommand{\assignment}{\Pi}

\newcommand{\hascon}{\footnotesize\texttt{connector}}

\newcommand{\elementuniverse}{\mathcal{E}}
\newcommand{\connectoruniverse}{\mathcal{C}}
\newcommand{\assetuniverse}{\mathcal{A}}
\newcommand{\boundaryuniverse}{\mathcal{B}}
\newcommand{\itemuniverse}{\mathcal{I}}

\newcommand{\metamodel}{\mathcal{M}}

\newcommand{\itemvars}{X}
\newcommand{\itemvar}{x}
\newcommand{\projection}{\Pi}

\newcommand{\boundaryrelation}{\beta}
\newcommand{\assetrelation}{\alpha}

\newcommand{\src}{\textbf{s}}
\newcommand{\tgt}{\textbf{t}}

\newcommand{\holds}{{\footnotesize\texttt{holds}}}

\newcommand{\attval}{{\footnotesize\texttt{val}}}
\newcommand{\ttype}{{\footnotesize\texttt{type}}}

\newcommand{\pathelms}{\text{elements}}
\newcommand{\pathcons}{\text{connectors}}
\newcommand{\elementstart}{e_{\emph{start}}}
\newcommand{\elementend}{e_{\emph{end}}}

\newcommand{\pathvars}{P}
\newcommand{\pathvar}{p}
\newcommand{\isin}{\footnotesize\texttt{ in }}

\newcommand{\change}[1]{{\color{black}#1}}
\newcommand{\changenew}[1]{{\color{black}#1}}

\DeclareMathOperator*{\argmin}{arg\,min}

\DeclareMathOperator*{\solver}{solver}
\DeclareMathOperator*{\push}{push}
\DeclareMathOperator*{\pop}{pop}
\DeclareMathOperator*{\add}{add}
\DeclareMathOperator*{\addsoft}{add\,soft}
\DeclareMathOperator*{\solve}{solve}
\DeclareMathOperator*{\maxsolve}{max\,solve}
\DeclareMathOperator*{\model}{model}
\DeclareMathOperator*{\solverpush}{\solver.push}
\DeclareMathOperator*{\solverpop}{\solver.pop}
\DeclareMathOperator*{\solveradd}{\solver.add}
\DeclareMathOperator*{\solveraddsoft}{\solver.add\,soft}
\DeclareMathOperator*{\solversolve}{\solver.solve}
\DeclareMathOperator*{\solvermaxsolve}{\solver.max\,solve}
\DeclareMathOperator*{\solvermodel}{\solver.model}

\DeclareMathOperator*{\sat}{sat}
\DeclareMathOperator*{\unsat}{unsat}
\DeclareMathOperator*{\unknown}{unknown}

\DeclareMathOperator*{\repair}{repair}
\DeclareMathOperator*{\partialrepair}{PartialRepair}
\DeclareMathOperator*{\HPartialRepair}{HPartialRepair}
\DeclareMathOperator*{\SMTMaxSolver}{SMTMaxSolver}
\DeclareMathOperator*{\SMTSolver}{SMTSolver}
\DeclareMathOperator*{\hasattr}{has\,attr}

\newcommand{\revision}[1]{#1}
\usepackage{xspace}
\makeatletter
\DeclareRobustCommand\onedot{\futurelet\@let@token\@onedot}
\def\@onedot{\ifx\@let@token.\else.\null\fi\xspace}

\makeatother

\def\BibTeX{{\rm B\kern-.05em{\sc i\kern-.025em b}\kern-.08em
    T\kern-.1667em\lower.7ex\hbox{E}\kern-.125emX}}

\newtheoremstyle{definition}
{3pt}
{3pt}
{}
{}
{\bfseries}
{.}
{.5em}
{}

\lstdefinelanguage{SMT2}{
    keywords={exists,forall,let,and,weight,and,or,not},
    keywords=[2]{state,attr,type},
    keywords=[3]{StateId},
    keywordstyle={\color{blue}\bfseries},
    keywordstyle=[2]{\color{red!50!blue}},
    keywordstyle=[3]{\color{teal}},
    sensitive=false, 
    morecomment=[l]{\#}, 
    showstringspaces=false,
} %

\title{Threat Repair with Optimization Modulo Theories}

\author{Thorsten Tarrach\inst{1} \and Masoud Ebrahimi\inst{2} \and Sandra König\inst{1} \and Christoph Schmittner\inst{1} \and Roderick Bloem\inst{2} \and Dejan Nickovic\inst{1}}
\institute{AIT Austrian Institute of Technology \and Graz University of Technology}

\begin{document}
\maketitle

\begin{abstract}

We propose a model-based procedure for automatically 
preventing security threats using formal models. 
We encode 
system models and 
potential threats as satisfiability modulo theory (SMT) formulas. 
This model allows us to ask security questions as satisfiability queries. 
We formulate threat prevention as an optimization problem over the same formulas. The outcome of our threat prevention procedure is 
a suggestion of model attribute repair that eliminates threats. Whenever threat prevention fails, 
we automatically explain why the threat happens. We implement our approach using the state-of-the-art Z3 SMT solver and 
interface it with the threat analysis tool THREATGET. 
We demonstrate the value of our procedure in two case studies 
from automotive and smart home domains, including an industrial-strength example.
\end{abstract}

\section{Introduction}
\label{sec:intro}

The proliferation of communication-based technologies requires 
engineers to have 
cybersecurity in mind when designing new applications. 
Historically, security decisions in the early stages of development have been made informally.
\revision{The upcoming requirements 
regarding security compliance in soon-to-be-mandatory standards such as the  ISO/SAE 21434 call for 
more principled security assessment of designs}
\revision{and the 
need for systematic reasoning about system security properties has resulted 
in threat modeling and analysis tools. 
One example of this 
new perspective is the Microsoft Threat Modelling Tool (MTMT)~\cite{mcree2014microsoft}, 
developed as part of the Security Development Lifecycle.}
MTMT provides capabilities for visual system structure modelling. Another example is 
THREATGET~\cite{DBLP:conf/safecomp/SadanySK19,DBLP:journals/corr/abs-2107-09986}, a 
threat analysis and risk management tool, originally developed in 
academia and following its success, commercialized and used today by leading word-wide companies in automotive 
and Internet-of-Things (IoT) domains.
Threat modeling and analysis significantly reduces the difficulty of a security assessment, reducing it to accurate modeling of the systems and the security requirements. 

Existing methods use ad-hoc methods to reason about the security of systems. As a result, it is not easy to extend such tools with capabilities like 
automated threat prevention 
and model repair. Although a trial-and-error method is always possible, it does not provide a systematic exploration of the space of possible prevention measures and leaves the question of optimizing the cost of the prevention to the designer's intuition. As a result, remedying a potential threat remains cumbersome and simple solutions may be missed, especially when multiple interacting threats exist.

This paper proposes a procedure for preventing threats based on a formal model of the structure of the system and a logic-based language for specifying threats. The use of rigorous, formal languages to model the system and specify threats allows us to automate threat prevention. More specifically, we reduce the problem of checking presence of threats in the system model to a satisfiability modulo theory (SMT) check. A threat specification defines a class of potential threats and a witness of a system model that satisfies a threat specification defines a concrete threat in the model.
This allows us to frame the problem of preventing concrete threats as 
an attribute parameter repair. 

The attributes 
of system elements and communication links define 
a large spectrum of security settings and, in presence of a threat, of possible preventive actions. 
This class of repairs enables simple 
and localized measures whose cost can easily be assessed by a designer. We formulate attribute repair 
as a weighted maximum satisfiability (MaxSAT) problem with a model of cost of individual changes to the system attributes. This formulation of the problem allows us to find changes in the model with {\em minimal} cost that result in removing as many threats as possible. 

We introduce {\em threat logic} as a specification language 
to specify threats. 
We formalize the system model as a logic formula that consists of a conjunction of sub-formulas, called {\em assertions}, parameterized by attributes that specify security choices.  The conjunction of the system model formula and a threat formula is satisfiable iff that threat is present in the system.
We introduce clauses that \change{change the specific instantiation of} model attributes to \change{a different} value and associate a {\em weight} with each such assertion. Then, the MaxSAT solution of this formula is the set of changes to system model attributes with minimum cost that ensure the absence of the threat. Given an incorrect system, we can choose the weights so that we compute the set of changes to system model attributes with the minimal 
cost to remove the existing threats from the model. 
To ensure that our method scales to industrial size models, we also define a heuristic that provides partial threat prevention by addressing repairable threats and explaining the reason why the others cannot be repaired. We believe that this method, even though partial and approximate in general, can compute near optimal repairs for many real-world problems. 

We implemented the threat prevention method in the 
THREATGET tool and evaluated it on two case studies from the automotive and the IoT domains. 

\subsubsection{Motivating Example}
\label{sec:motivating}

\revision{We motivate this work with a smart home application 
from the IoT domain, depicted in 
Figure~\ref{fig:IoT}. 
%
The smart home architecture consists of $7$ 
\emph{typed elements}: (1) a control system, (2) 
an IoT field gateway, (3) temperature and 
(4) motion sensors, (5) a firewall, (6) 
a web server and (7) a mobile phone. The 
elements are interconnected using \emph{wired} 
and \emph{wireless connectors}. The 
elements and connectors have associated sets 
of \emph{attributes} that describe their 
configuration. 
For instance, every connector has attributes Encryption, Authentication and Authorization. The attribute Encryption can be assigned the values No, Yes and Strong. We associate to each attribute a {\em cost} 
of changing the attribute value, reflecting our assessment of how difficult it is to implement the change.
In this example, the temperature and the motion sensor communicate wirelessly with the gateway. If the motion sensor detects a movement, the user is notified by phone. It is possible to override the behavior, e.g., the heating can be turned on remotely in case of late arrival.
The web server protected by the firewall allows for access and information exchange from and to the smart home. The IoT sub-system protected by the firewall defines a \emph{security boundary} called 
the IoT Device Zone. 
Communication should be confidential and encrypted outside the IoT Device, which is represented by the two associated \emph{assets}.} 
\revision{
\vspace{-12pt}
\begin{figure}
  \centering
  \resizebox{\columnwidth}{!}{\input{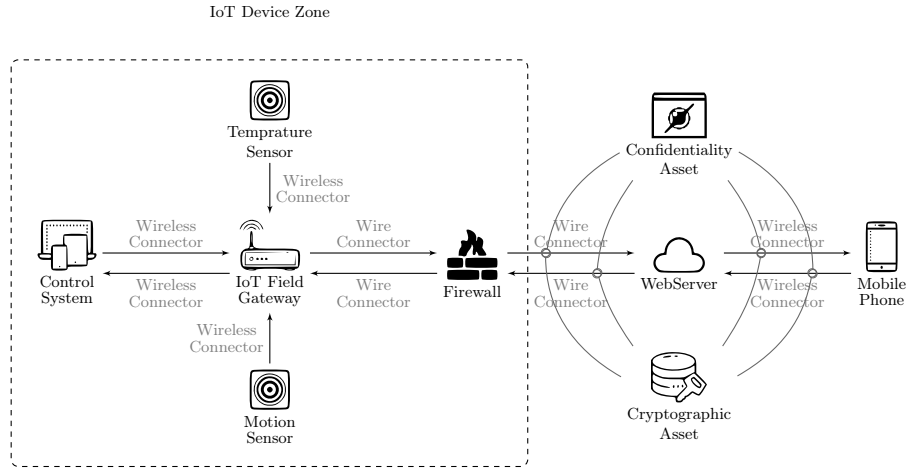}}
 \caption{Smart Home IoT model.}\label{fig:IoT}
 \vspace{-12pt}
\end{figure}
}

\revision{The potential threats that can be 
encountered in this smart home system 
are characterized by logical relations between elements, connectors 
and their attributes. Consider the two potential threats that are applicable to this example:
\begin{description}
\item[Threat 1:] The web server enables data logging functionality without encrypting the data.
\item[Threat 2:] The mobile phone device is  connected to the web server, 
without the web server enabling data logging.
\end{description}

Assume that the web server has data logging enabled, but no data encryption, thus matching Threat 1. If we consider this threat in isolation, we can repair it in two ways: 
(1) by turning off the data logging, or (2) by implementing the data encryption on the web server. 
The first repair results in matching Threat 2. Only the second repair results in the removal of all security threats. Given two data encryption algorithms with costs $c_1$ and $c_2$, where $c_1 > c_2$, implementing 
the latter is
the cost-optimal option. We see that an optimal preventive solution must consider simultaneous repair of multiple threats.}

\section{Threat Modelling}
\label{sec:model}

A threat model consists of two main components, 
a \emph{system model} and a database of 
\emph{threat rules}. A system model provides an architectural view of the system under investigation, representing relevant components, their properties, as well as relations and dependencies between them
\footnote{The system and the threat model are formally defined in
Appendices~\ref{appendix:model} and~\ref{appendix:logic}.
}.

\subsubsection{System Model}
\label{sec:system}


\revision{
A system model $\threatmodel$ consists of:
\begin{itemize}
    \item a set $\elements$ of \emph{elements}: an element 
    $e \in \elements$ is a typed 
    logical (software, database, etc.) or physical (ECUs, sensors, 
    actuators, etc.) component.
    \item a set $\connectors$ of \emph{connectors}: a connector 
    $c \in \connectors$ is a direct 
    interaction between two elements, a \emph{source} 
    $\src(c) \in \elements$ and a \emph{target} element $\tgt(c) \in \elements$.
    \item a set $\assets$ of \emph{security assets}: an asset
    $a \in \assets$ describes logical or physical object (such as an 
    element or a connector) of value. Each element and connector can hold multiple assets. Similarly, each asset can be associated to multiple elements and connectors.
    \item a set $\boundaries$ of \emph{security boundaries}: 
    a boundary $b \in \boundaries$ describes a separation between logically, physically, or legally separated system elements. 
    \item a set $\attributeuniverse$ of \emph{attributes}: an attribute $\attribute \in \attributeuniverse$ 
    is a property that 
    can be associated to a system elements, connectors and/or assets. 
    Each attribute $\attribute$ can assume a value from its associated domain $D_\attribute$. We 
    denote by $\attributevalueassignment(x, \attribute)$ the value 
    of the attribute $\attribute$ associated to the 
    element/connector/asset $x$. We finally define an 
    \emph{attribute cost} mapping 
    $\attributeweight_{x,\attribute}(v,v')$ associated to 
    $(x,\attribute)$ pairs that defines the cost of changing 
    the attribute value $v \in D_\attribute$ to $v' \in D_\attribute$.
\end{itemize}

Given a system model $\threatmodel$, we define a \emph{path} 
$\pi$ in $\threatmodel$ 
as an alternating sequence $e_{1}, c_{1}, e_{2}, c_{2} \cdots, c_{n-1},e_{n}$ of elements and connectors, 
such that for all $1 \leq i \leq n$, $e_{i} \in E$, for all $1 \leq i  < n$, $c_{i} \in C$, $\src(c_{i}) = e_i$, and $\tgt(c_{i}) = e_{i+1}$ and for all $1 \leq i < j \leq n$, $e_{i} \neq e_{j}$. We note that we define paths to 
be \emph{acyclic}, since acyclic paths are sufficient to express 
all interesting security threats.

We use the notation $\pathelms(\pi)$ and $\pathcons(\pi)$ to 
define the sets of all elements and of all connectors appearing in a path, respectively. The starting and the ending element in the path $\pi$ are denoted by $\elementstart(\pi) = \src(c_{1})$ and $\elementend(\pi) = \tgt(c_{n-1})$, 
respectively. 
We denote by $P(\threatmodel)$ the set of all paths in $\threatmodel$.



} 
\subsubsection{Threat Logic}
\label{sec:logic}

\revision{
We provide an intuitive 
introduction of 
{\em threat logic} for specifying potential threats
\footnote{The authors of THREATGET define their own syntax and semantics to express threats \cite{DBLP:journals/corr/abs-2107-09986}. We use instead predicate logic to facilitate the 
encoding of the forthcoming algorithms into 
SMT formulas. 
Our implementation contains an automated 
translation from THREATGET syntax to threat logic.}.
%
The syntax of threat logic is defined as follows:
\begin{equation*}
\begin{array}{lcl}
\varphi &  := & R(\itemvars \cup \pathvars)~|~\neg \varphi~|~\varphi_1 \vee \varphi_2~|~\exists \pathvar. \varphi~|~\exists \itemvar. \varphi \\
\end{array}
\end{equation*}
\noindent where $X = \elements \cup \connectors 
\cup \assets \cup \boundaries$, $\itemvar \in \itemvars$, 
$\pathvars$ is a set of path variables, 
$\pathvar \in \pathvars$, and 
$R(\itemvars \cup \pathvars)$ is a predicate. 
The predicate $R(\itemvars \cup \pathvars)$ 
is of the form:
\begin{enumerate}
    \item $\ttype(x) = t$ - the type of $x \in \itemvars$ is $t$;
    \item $x \isin p$ - the element or the connector $x \in \elements \cup \connectors$ is in the path $p \in \pathvars$;
    \item $\hascon(e, c)$ - the element $e \in \elements$ is either the source or the target of the connector $c \in \connectors$;
    \item $\source(c) = e$ - the 
    source of the connector $c \in \connectors$ is the element $e \in \elements$;
    \item $\target(c) = e$ - the target of the connector $c \in \connectors$ is the element $e \in \elements$;
    \item $\source(p) = e$ - the 
    source of the path $p \in \pathvars$ is the element $e \in \elements$;
    \item $\target(p) = e$ - the target of the path $p \in \pathvars$ is the element $e \in \elements$;
    \item $\crosses(c, b)$ -  the connector $c \in \connectors$ crosses the boundary $b \in \boundaries$;
    \item $\contains(x, b)$ - the element or boundary $x \in \elements \cup \boundaries$ is contained in the boundary $b \in \boundaries$; 
    \item $\holds(x, a)$ - the element or the connector $x \in \elements \cup \connectors$ holds the asset $a \in \assets$;  \item $\attval(x, \emph{att}) = v$ -  the valuation of the attribute {\em att} 
    associated to $x \in \elements \cup \connectors \cup \assets$ is equal to $v$.
\end{enumerate}

\begin{example}
Consider a requirement that there exists a  
path in the model such that all the elements in that path  are of type Cloud. It is expressed with the threat logic formula: $
\exists p.\forall e.(e \isin p \implies \ttype(e)= \text{Cloud}).
$
\end{example}

We define an \textit{assignment} 
$\projection_{\threatmodel}$ as a partial function 
that assigns element, connector, asset, 
security boundary and path variables to concrete 
elements, connectors, assets, security 
boundaries and paths from the system 
architecture model $\threatmodel$.
We denote by $\projection_{\threatmodel}[\itemvar \mapsto i]$
the item assignment in which $\itemvar$ is mapped to $i$ and otherwise identical to \change{$\projection_{\threatmodel}$}.
Similarly, we denote 
by $\projection_{\threatmodel}[\pathvar \mapsto \pi]$
the path assignment in which $\pathvar$ is mapped to $\pi$ and otherwise identical to \change{$\projection_{\threatmodel}$}.
The semantics of threat logic follow the 
usual definitions of predicate logic. 

We say that a threat logic formula is {\em closed} when all occurrences of element, connector, 
asset and security boundary variables 
are in the scope of a quantifier. Any closed threat logic formula is a valid threat 
specification. Given a system model $\threatmodel$ and a closed threat logic 
formula $\varphi$, we say that $\threatmodel$ {\em witnesses} the threat $\varphi$, 
denoted by $\threatmodel \models \varphi$ iff $\assignment_{\threatmodel} \models \varphi$, 
where $\assignment_{\threatmodel}$ is an empty assignment.

\subsubsection*{From Threat Logic To First Order Logic (FOL)}

We observe that we interpret threat logic formulas over system models with a finite number of elements and connectors, and hence we can eliminate path quantifiers by enumerating the elements and connectors in the path. By eliminating path quantifiers, we obtain an equisatisfiable 
FOL formula that can be directly used 
by an SMT solver. 
The path quantifier elimination procedure $T$ that takes as input a threat 
formula $\varphi$ and computes the equisatisfiable
first-order formula $T(\varphi)$
is formalized in Appendix~\ref{tab:sem:appendix}.

\begin{example}
\label{ex:rules}
We formalize the two threats described in Section~\ref{sec:motivating}:

\footnotesize
    \begin{tabular}{ll}
    Threat 1     &  $\exists e. (\ttype(e) = \text{WebServer} \wedge \attval(e, \text{Data Logging}) = \text{Yes})$ \\
                &  $\wedge \attval(e, \text{Data Encryption}) \neq \text{Yes})$   \\
    
    Threat 2     &  \change{$\exists p, e_1, e_2. 
    \source(p) = e_1 \wedge \target(p) = e_2 \wedge$} \\
    &
    \change{$\ttype(e_2) = \text{WebServer} \; \wedge \ttype(e_1) = \text{MobilePhone} $} \\
                & \change{$\wedge \attval(e_2, \text{Data Logging}) \neq \text{Yes})$} \\
    \end{tabular}

\end{example}
}

\section{Automated Threat Prevention}
\label{sec:mitigation}

We now present our main contribution -- a procedure to automatically repair a \change{system} model with one or more threats. 
We restrict ourselves to the class of {\em attribute repairs} that consists in changing the model attribute values (see Section~\ref{sec:attribute}). 
We show how to encode the attribute repair problem 
as an optimization problem modulo theories in Section~\ref{sec:smt} 
and present an exact algorithm for doing the minimal attribute repair using an SMT solver. In Section~\ref{sec:unrepairable}, 
we address the problem of partial repair in presence of 
a unrepairable threats. Finally, we propose in Section~\ref{sec:heuristic} a more scalable and flexible heuristic for partial repair.

\subsection{Attribute Repair}
\label{sec:attribute}

In this work, repairing a model that has one or multiple 
threats consists in changing the attribute valuation function of the model. Not every model can be attribute repaired. 
\changenew{For a threat model $\threatmodel$ we denote by $\threatmodel[\attributevalueassignment' \backslash \attributevalueassignment]$ the threat model in which the attribute 
value assignment $v$ is replaced by another 
assignment $v'$.}

\begin{definition}[Threat-repairable model]
\label{def:repairable}
Given a model $\threatmodel$ with an attribute valuation function 
$\attributevalueassignment$ that witnesses a threat $\varphi$,  
$\threatmodel \models \varphi$, we say that 
$\threatmodel$ is {\em attribute repairable} wrt $\varphi$ 
iff there exists a $\attributevalueassignment'$ such that 
$\threatmodel[\attributevalueassignment' \backslash \attributevalueassignment] \not \models \varphi$. 
\end{definition}

We specifically aim at finding the {\em optimal} 
repair, which has the minimal repair cost.
To reason about this quantitative 
repair objective, Definition~\ref{def:distance} specifies the {\em distance}  
$d(\attributevalueassignment, \attributevalueassignment\change{'})$ 
between two attribute valuation functions $\attributevalueassignment$  and $\attributevalueassignment\change{'}$ as the sum of altered attribute costs for attributes
that differ in the two valuations.


\begin{definition}[Attribute valuation distance]
\label{def:distance}
Let $\threatmodel$ be a system model with attribute valuation  
$\attributevalueassignment$ and attribute cost 
$\attributeweight$. Let $\attributevalueassignment'$ be another 
attribute valuation.
The {\em distance} 
$d(\attributevalueassignment, \attributevalueassignment')$ between $\attributevalueassignment$ and 
$\attributevalueassignment'$ is defined as:
\change
{
\[
d(\attributevalueassignment,\attributevalueassignment') =
\sum\limits_{x \in X, \attribute \in \attributeuniverse} \attributeweight_{x, \attribute}(\attributevalueassignment(i, \attribute),\attributevalueassignment'(x, \attribute))  \text{~s.t.~} \attributevalueassignment(x, \attribute) \neq 
\attributevalueassignment'(x, \attribute).
\]
}
\change{
For instance, for the attribute `Encryption' the cost of changing from `None' to `Weak' may be 20, but to change `None' to `Strong' may cost 30. A change from `Weak' to `Strong' could cost 15, but a change from `Weak' to `None' may only cost 1.
Sensible cost functions will adhere to some restrictions (such as a variant of the triangle inequality) that we do not formalize here.}

\end{definition}


\begin{definition}[Minimal attribute repair]\label{def:mitigation}
Let $\varphi$ be a threat logic formula and $\threatmodel$ a 
system model such that 
$\threatmodel \models \varphi$ and $\threatmodel$ is attribute 
repairable w.r.t. $\varphi$. 
The {\em minimal attribute repair} of $\threatmodel$ is another 
threat model $\threatmodel[v'\backslash v]$ 
s.t.:
\[
\attributevalueassignment' = \argmin \{d(\attributevalueassignment, \attributevalueassignment'') | \threatmodel[\attributevalueassignment''\backslash\attributevalueassignment]  \not \models \varphi\}
\]
\end{definition}

\noindent \emph{Other notions of minimal repair.}
There are at least two other natural notions of minimal repair. 
In the first one,
\change{costs are associated with the attribute itself. This means that every change of the attribute carries the same cost.
We can model this by assigning the same cost to all possible combinations of previous and new value for an attribute}.
Alternatively, 
engineers are often not interested in 
minimizing the overall real cost, but rather 
in minimizing the number of attributes that need 
to be repaired. We can 
model this restricted variant of the problem 
by associating the fixed cost of $1$  each 
attribute in the model, thus effectively counting the number of individual attribute repairs.
Both variants can be implemented in a straightforward manner in 
our framework.

\subsection{Attribute Repair as Weighted MaxSMT}
\label{sec:smt}

We encode the attribute 
repair problem (see Section~\ref{sec:attribute}) as 
a {\em weighted MaxSMT} problem, in which 
$\mathbf{F}$ represents a (hard) {\em assertion}, while 
$F_{1}, \ldots, F_{\change{m}}$ correspond to {\em soft assertions} and 
every soft assertion $F_i$ has an associated cost $\mathit{cost}_i$. 

\begin{definition}[Weighted MaxSMT~\cite{DBLP:conf/sycss/BjornerP14}]
Given an SMT formula $\mathbf{F}$, a set of SMT formulas $F_{1}, \ldots, F_{m}$ 
and a set of real-valued costs $\mathit{cost}_{1}, \ldots, \mathit{cost}_{m}$, 
the {\em weighted MaxSMT} problem consists in finding a 
subset $K \subseteq M$ of indices with $M = \{1, \ldots, m\}$ such that: (1) $\mathbf{F} \wedge \bigwedge_{k \in K} F_{k}$ is satisfiable, and (2) the total cost $\change{\sum}_{\change{i} \in N\backslash K} \mathit{cost}_{\change{i}}$ is minimized.
\end{definition}


\begin{table}
    \vspace{-30pt}
    \centering
    \caption{Methods in SMTMax solver.}
    \label{tab:smtmaxsolver}
    \begin{tabular}{ll}
    \toprule 
    \textbf{Method}     &  \textbf{Description} \\
    \midrule
    $\push()$  & Push new context to  solver stack \\
    $\pop()$   & Pop context from  solver stack \\
    $\add(\varphi)$   & Add new hard assertion $\varphi$ \\
    $\addsoft(\varphi, \mathit{cost})$ & Add new soft assertion $\varphi$ with weight \change{$\mathit{cost}$}\\
    $\solve()$ & Check if  formula is satisfiable \\
    $\maxsolve()$ & Check if  formula is max-satisfiable \\
    $\model()$ & Generate and return a model witnessing  satisfaction of formula \\
    \bottomrule
    \end{tabular}
\end{table}

We now sketch the encoding of the minimal attribute repair problem 
into weighted MaxSMT. We assume that we have a MaxSMT solver 
object, with the functionality described in Table~\ref{tab:smtmaxsolver}. 


Given a system model $\threatmodel$ and a set of threat logic formulas 
$\Phi = \{ \varphi_1, \ldots, \varphi_n \}$, we compute the MaxSMT formulas 
$\textbf{F}$ that represents
the hard assertion $\mathbf{F} = F_\threatmodel \land \bigwedge_{j=1}^{n} \neg \varphi_j$
conjoins $F_{\threatmodel}$ that encodes the entire system model \textit{except} its attribute valuations and costs with the negation of each threat logic formula $\varphi_{j}$.
We also define one soft assertion $F_{x, \attribute, v}$ for each element $x$, attribute $\attribute$ of $x$ and possible value $v$ of $\attribute$, stating intuitively that $\attributevalueassignment(x, \attribute) = v$. These soft attributes are mutually exclusive if they assert different values for the same attribute. We set up the costs of each $F_{x, \attribute, v}$ in such a way that asserting $F_{x, \attribute, v}$ leads to cost corresponding to changing the value of $\attribute$ to $v$. (The exact value of the cost function can easily be computed by solving a linear system of equations.)





We use the weighted MaxSAT solver 
$\maxsolve(F_{\threatmodel} \wedge \bigwedge_{j=1}^{n} \neg \varphi_{j} \wedge \bigwedge F_{x, \attribute, v})$
to 
obtain the satisfiability verdict and the optimization cost. 
Informally, the solver can return three possible verdicts:\footnote{We ignore here 
a fourth possible verdict $\unknown$ that can arise in practice and that happens 
if the solver is not able to reach a conclusion before it times out.}
\begin{itemize}
    \item $\sat$ verdict with total cost $0$: 
    the system model $\threatmodel$ does not contain a potential threat defined by 
    any of the threat formulas $\varphi_i$,
    \item $\sat$ verdict with total cost $k$: the system model $\threatmodel$ contains a set of potential threats defined by a subset of threat formulas and can be repaired by changing the values of model attributes with total cost $k$. The solver returns a model, which defines a possible repair, i.e. the altered attribute values that render the formula satisfiable,
    \item $\unsat$ verdict: the system model $\threatmodel$ is not attribute repairable with respect to at least one threat formula $\varphi_i$.
\end{itemize}

The encoding of the attribute repair into this MaxSMT problem provides 
an effective solution to the minimum attribute repair problem. 

\begin{restatable}{theorem}{thmmain}
\label{th:main}
Let $\threatmodel$ be a system model and $\{\varphi_1, \ldots, \varphi_{n}\}$ a set of closed threat logic formula. We have that $\maxsolve(F_{\threatmodel} \wedge \bigwedge_{i=1}^{n} \neg \varphi_{i} \wedge \bigwedge_{F \in \Psi} F)$ provides the solution to the minimum attribute repair problem.
\end{restatable}

\subsection{Partial Repair of Unrepairable Models}
\label{sec:unrepairable}

The problem with the approach from Section~\ref{sec:smt} 
arises if there is a formula 
$\varphi_{i}$ for which $\threatmodel$ is not 
attribute repairable. In that 
case, the entire problem is unsatisfiable, even if other threats 
could be repaired. This outcome, although correct, is not 
of particular value to the security engineer. Ideally, 
the objective is to repair attributes for threats that 
can be repaired and explain the others.

We observe that attribute-unrepairable threats have a particular form and correspond to  formulas without constraints 
on attribute valuations. An inductive visit of the formula allows a syntactic check $\text{has\_attr}(\varphi)$ whether a threat formula $\varphi$ has any constraint on attribute valuations. 
Algorithm~\ref{alg:full} implements $\partialrepair$, a method that adapts the MaxSMT algorithm from Section~\ref{sec:smt} 
to compute a partial repair 
of $\threatmodel$ with respect to a subset of repairable threat formulas. The procedure removes all threat logic formulas 
that are satisfied by the model and that are known to be 
unrepairable, before computing MaxSMT.

From the definition of the partial repair algorithm, it  
follows that the MaxSAT applied to the subset of (potentially) 
repairable threat formulas corresponds to the minimum 
attribute repair restricted to that subset of threat formulas.
\begin{corollary}
\label{prop:restricted}
Let $\threatmodel$ be a system model, $\Phi = \{\varphi_1, \ldots, \varphi_{n}\}$ a set of closed threat logic formulas and 
$G \subseteq \Phi$ a subset of \change{repairable threats, 
i.e. for all $\varphi \in G$, $M \not\models \varphi$ or 
$\hasattr(\varphi)$ is true.}
We have that \change{$\maxsolve((F_{\threatmodel} \wedge \bigwedge_{\varphi \in  G} \neg \varphi) \wedge \bigwedge_{F \in \Psi} F)$} provides the solution to the minimum attribute repair problem restricted to the set \change{$G$} 
of threat formulas.
\end{corollary}

\begin{algorithm2e}[H]
\SetKwInOut{Input}{Input}
\SetKwInOut{Output}{Output}
\Input{$\threatmodel$, $\{ \varphi_{1}, \ldots, \varphi_{n}\}, F_\threatmodel, \hat{F}_\threatmodel, \Psi $}
\Output{status of repair, cost of repair,\\list of repaired attribute values $\hat{\Psi}$}
$\solver \gets \SMTSolver()$ \;
$\solveradd(\hat{F}_\threatmodel)$ \;
$G \leftarrow \emptyset$ \;


\For{$\varphi \in \{ \varphi_{1}, \ldots, \varphi_{n} \}$}{
    $\solverpush()$ \;
    $\solveradd(\changenew{T(\varphi)})$ \;
    $status \gets \solversolve()$ \;
    $\solverpop()$ \;
    \If{$status = \sat$}{
        \If{$\hasattr(\varphi)$}{
            $G \gets G \cup \{ \varphi \}$ \;
        }
    }
    \ElseIf{$status = \unsat$}{
        $G \gets G \cup \{ \varphi \}$ \;
    }
}


$\solver \gets \SMTMaxSolver()$ \;
$\solveradd(F_\threatmodel)$ \;

\For{$F \in \Psi$}{
    $\solveraddsoft(F,\mathit{cost}(F))$ \;
}

\For{$\varphi \in G$}{
    $\solveradd(\neg \changenew{T(\varphi)})$ \;
}

$status \gets \solvermaxsolve()$ \;
$\hat{\Psi} \leftarrow \emptyset$ \;
$totalcost \gets 0$ \;

\If{$status = \sat$}{
    $ m \gets \solvermodel()$ \;
    $\hat{\Psi}, totalcost \gets \repair(m, \Psi)$ \;
}

    \Return $status, totalcost, \hat{\Psi}$
\caption{Partial attribute repair: $\partialrepair()$\label{alg:full}}
\end{algorithm2e}


\noindent \emph{Explaining unrepairable threats:} The partial 
repair method is useful in the presence 
of threats that cannot be addressed by attribute repair only. 
The SMT solver can be used to provide in addition an explanation of 
why a threat cannot be repaired -- 
the solver assigns values to variables in threat logic formulas 
that witness its satisfaction (i.e. the presence of that threat). 
This witness explains exactly why that formula is satisfied and 
locates in the system model the \change{one set of} items that are responsible for that verdict. \change{The threat may match at multiple locations of the model, which could be discovered by multiple invocations of the solver (while excluding the previously found sets). Note that the MaxSMT will repair all occurrences though, because the threat is negated there and the negated existential quantifier becomes a forall quantifier.}


\subsubsection{Heuristic for Partial Repair}
\label{sec:heuristic}

Removing unrepairable formulas in Algorithm~\ref{alg:full} 
does not guarantee that the remaining  
threat formulas can be repaired together. This can happen 
if for example there is a pair of threat logic formulas that 
are mutually inconsistent with respect to the system model.
We propose a heuristic procedure for partial repair in Algorithm~\ref{alg:approximate}.

\begin{algorithm2e}[H]
\SetKwInOut{Input}{Input}
\SetKwInOut{Output}{Output}
\Input{$\threatmodel$, $\{ \varphi_{1}, \ldots, \varphi_{n} \}, F_\threatmodel, \Psi$}
\Output{Repair status and cost, set of repaired and non-repaired threats, repaired set of attributes}
$nothreat \gets \emptyset\,;~repairable \gets \emptyset\,;~
totalcost \gets 0$ \;

$\solver \gets \SMTMaxSolver()$ \;
$\solveradd(F_\threatmodel)$ \;

\For{$\varphi \in \{ \varphi_{1}, \ldots, \varphi_{n} \}$}{
    $\solverpush()$ \;
    \For{$F \in \Psi$}{
        $\solveradd(F)$ \;
    }
    $\solveradd(\changenew{T(\varphi)})$ \;
    $status \gets \solversolve()$ \;
    $\solverpop()$ \;
    \If{$status = \unsat$}{
        $nothreat \gets nothreat \cup \{ \varphi \}$ \;
    }
    \ElseIf{$status = \sat$}{
        \If{$\hasattr(\varphi)$}{
            $\solverpush()$ \;
            \For{$F \in \Psi$}{
              $\solveraddsoft(F, \mathit{cost}(F))$ \;
            }
            $\solveradd(\neg \changenew{T(\varphi)})$ \;
            $status \gets \solvermaxsolve()$ \;
            \If{$status = \sat$}{
                $m = \solvermodel()$ \;
                $\hat{\Psi}, c \gets \repair(m, \Psi)$ \;
                $\solverpop()$ \;
                $\solverpush()$ \;
                $\changenew{\solveradd(\bigvee_{\varphi' \in repairable \cup nothreat}T(\varphi'))}$ \;
                \For{$F \in \hat{\Psi}$}{
                  $\solveradd(F)$ \;
                }
                $status \gets \solversolve()$ \;
                \If{$status = \unsat$}{
                    $repairable \gets repairable \cup \{ \varphi \}$ \;
                    $\Psi \gets \hat{\Psi}$ \label{line:repair} \;
                    $totalcost \gets totalcost + c$ \;
                }
            } 
            $\solverpop()$ \;
        }
    } 
}

\Return $status, totalcost, repairable, nothreat, \Psi$
\caption{\small Approximate~partial~repair:~$\HPartialRepair$\label{alg:approximate}}
\end{algorithm2e}

The procedure takes as input the solver (we assume that 
it already has the system model encoding that includes 
all hard assertions), the set of attribute valuations given 
in the form of assertions, and the set of threat  logic formulas.
The procedure maintains two sets of threat logic formulas, 
the ones that do not pose any threat to the system model 
and the ones that do pose a threat, but are repairable (line 1). The 
set of unrepairable formulas are implicitly the ones that 
are in neither of these two sets.

For every threat logic formula $\varphi$, the procedure 
first checks if that threat is present in the system model 
using the SMT solver (lines 5--10). If the threat is absent, 
it is added to the set of formulas that do not represent 
any threat (lines 11--12). Otherwise, the procedure attempts 
to compute a repair for that particular threat (lines 13--33). 
It first checks whether the threat logic formula refers to 
any attribute valuations (line 14). If not, the formula is 
unrepairable. Otherwise (lines 15--19), the  formula 
is added as a hard assertion, and the set of 
attribute valuations are added as soft assertions 
are added as soft assertions to the solver, and the MaxSMT 
solver is invoked. If the solver gives $\unsat$ verdict, it means that 
the model cannot be repaired to satisfy the threat logic formula. 
Otherwise (lines 20--32), we use the model witnessing the satisfaction 
of the threat logic formula to compute the partial repair 
for that formula (line 22). The outcome of the repair method 
is a the repaired set of attribute assertions and the number 
of assertions that needed to be altered. 
The procedure checks that this partial 
repair is consistent with the previous repairs (i.e. that it 
does not lead to violation of other previously processed threat logic 
formulas), and the repair is accepted only upon passing this last 
consistency check (lines 24-32).

\section{Implementation and Case Studies}
\label{sec:cs}

We implemented the methods presented in this paper in a prototype tool that we integrated to THREATGET threat analysis tool. 
THREATGET has a database of threat descriptions 
from automotive, IoT and other application domains against which the system model is analysed. The threat 
descriptions originate from multiple sources -- security-related standards, previously discovered threats, domain knowledge, 
etc, and can be extended with new threats. 

The threat prevention implementation imports system models as JSON files and threat descriptions as patterns written in the 
above-mentioned rule-based language. The tool translates both 
the model and the treat patterns into first-order logic SMT formulas and uses the Z3 solver for SMT and weighted MaxSMT. 
The outcomes of the MaxSMT solver are used to compute the repair suggestions. The implementation and the interface to Z3 are done in Java. THREATGET is a proprietary software (with a free academic license), the threat repair extension presented in this paper is distributed under the BSD-3 license. 
We observe that THREATGET 
does not currently support specification of attribute costs. \change{Hence, we allow the user to input the attribute change cost using a CSV file and assume default cost 1 for all attributes not mentioned in the CSV file.}

\revision{
We apply our tool to two case studies\footnote{A third key fob case study can be found in Appendix~\ref{sec:key:appendix}, in which we compare the vanilla MaxSMT approach 
to the heuristic procedure.} from two domains:  the smart home IoT application introduced as our motivating example in 
Section~\ref{sec:intro}, and  
the vehicular telematic gateway. }
%
We focus 
in each case study on a different aspect of our approach. 
The smart home IoT application 
is the only example where both the model and the set of 
threats are available in the public domain. Hence, in that 
case study, we illustrate various aspects of the model 
repair procedure on the concrete model and concrete threats.
The vehicular telematic gateway is an 
industrial-strength example developed with an industrial partner 
specializing in high-level control solutions. 
We use this case study to evaluate the scalability of 
our approach to real-world examples. 

\subsubsection{Smart Home IoT Application}

\revision{
This case study was introduced 
in Section~\ref{sec:intro}. In this section, 
we report on the experimental results obtained 
by applying our threat repair approach.}
The model has been analysed against 169 IoT-related threat descriptions given in the form of threat log formulas.
We applied both the full MaxSMT optimization procedure and its heuristic variant.

Both the model and the database 
of IoT threat formulas are publicly available.
Table~\ref{tab:iot} summarizes the outcomes.  \change{To accurately report the number of repairable formulas we implement Algorithm \ref{alg:approximate} without line \ref{line:repair}. The cost reflects the number of attributes (per item) changed in the model because we set the cost per attribute to 1 in our experiments.} We can observe that the heuristic procedure was able to repair 27 out of 36 found threats in less than 50 seconds.

\begin{table}[htb]
    \vspace{-20pt}
    \centering
    \caption{Results of attribute repair applied to Smart Home IoT case study.}
    \label{tab:iot}
    \begin{tabular}{|lr|lr|}
    \hline
    verdict     & SAT & \# formulas w/t threat &  \change{133} \\
    total \# formulas & 169 & total cost &  \change{77} \\
    \# repairable formulas & 27 & time (s) &  \change{46.7} \\
    \# unrepairable formulas &  9 &&\\ 
    \hline
    \end{tabular}
    \vspace{-10pt}
\end{table}

We illustrate the repair process on two threats 
from the database of IoT security threats. 
We first consider the threat with the title 
``Attacker can deny the malicious act and 
remove the attack foot prints leading to repudiation issues''. 
This threat is formalized using the  
threat logic formula
$$
\begin{array}{ll}
\exists e. & {\ttype}(e) = \text{Firewall}~\wedge  
(v(e, \text{Activity Logging}) \neq \text{Yes}~\vee 
v(e, \text{Activity Logging}) = \text{Missing}). \\
\end{array}
$$
\noindent This threat logic formula is translated into 
the SMT formula:
\begin{lstlisting}[
    basicstyle=\ttfamily\footnotesize,
    language=SMT2
]
(exists ((e1 StateId))
 (! (let ((a!1 (or (not 
  (= (state-attr e1 |Activity Logging|) Yes))
  (= (state-attr e1 |Activity Logging|) missing))))
  (and (= (state-type e1) Firewall) a!1)) :weight 0))
\end{lstlisting}

The SMT solver finds that the smart home IoT application
model satisfies this formula and hence has a threat. 
By extracting the witness from the solver, we find that 
the quantified variable $e$ is instantiated to the element 
with the identifier number $46$ of type `Firewall' and with 
the `Activity Logging' attribute set to `undefined'. This 
witness represents the explanation of the threat. The 
prevention algorithm proposes a repair that consists 
in setting the attribute `Activity Logging' to `Yes', i.e. 
implementing the activity logging functionality. This 
repair was found to be consistent with the other ones 
and is reported as part of the overall repair suggestion.

The second threat has the title ``Spoofing IP'' and 
is reported as an irreparable threat. It is formalized 
using the threat logic formula
$$
\begin{array}{ll}
\exists c. \exists e_1. \exists e_2 & \ttype(c) = \text{Internet Connection}~\land 
\source(c) = e_{1} \wedge \target(c) = e_{2}. \\
\end{array}
$$

The SMT solver identifies a satisfaction witness for this 
formula -- a connector of 
type 'Internet Connection' with id number 1 and its source and 
target element having id numbers 2 and 6. This threat cannot be repaired by changing the model attributes. On the contrary, this threat 
formula states that any connection to the internet constitutes 
a potential IP spoofing threat.

\subsubsection{Vehicular telematic gateway}

This study is based on an industrial-strength model of a vehicular telematic gateway (VTG)
\cite{TGCoop}. The VTG connect internal elements of the vehicle with external services. it offers vehicle configuration and entertainment and navigation to the user. An item is a term introduced by ISO 26262, describing a system or combination of systems, enabling a function on the vehicle level. 
The industrial company that devised the THREATGET model develops such systems for usage by different vehicle manufacturer so that the item is developed based on assumptions about the vehicle. For configuration and adaption in the target vehicle these assumptions need to be validated.

The case study considers a telematic ECU that offers remote connectivity for the on-board network to support various remote services including data acquisition, remote control, maintenance, and over-the-air (OTA) software update. In addition to that, it provides a human-machine interface (HMI) for navigation, configuration, and multimedia control. It represents the central element in a vehicle, connecting the control system to the human operator and the backend. It does not directly control the operation of the vehicle but is connected to such systems.
It has cellular and wireless local area network (WLAN) communication interfaces for wireless connectivity. For local connectivity, it includes a USB port for local software updates and application provision. The telematics system is also connected to other onboard ECUs.

For our analysis we considered two variants of time-triggered control. A simplified model contains 15 elements, 24 transitions and 5 assets, while the full model has 25 elements, 48 transitions and 5 assets.  
The presence of two models reflects the iterative design process in 
which the high-level simplified model was refined into the 
complete model based on the previous analysis.


A model of the full version is shown in Figure~\ref{fig:TTcontrol} 
(Appendix~\ref{sec:ttcontrol:appendix}). The results of the attribute repair are presented in Table~\ref{tab:ttcontrol}. We see that 
the tool is able to scale to large industrial models. It analyses \change{95} threat formulas in \change{497s}, repairing \change{19} out of \change{42} 
threats with the cost \change{57}. The same set of rules were 
repaired in \change{127s} on the simplified model. The main complexity in repairing large models comes from 
the threat formulas that contain quantification over path. 
This is not surprising because each such formula 
corresponds to a bounded model checking (reachability) 
problem. To confirm this observation, we analysed the full 
model without formulas with quantification over paths. 
The analysis of \change{82} such formulas was done in less than \change{118s}, 
resulting in the repair of \change{18} out of \change{39} threats.

\begin{table}
    \vspace{-25pt}
    \centering
    \caption{Results of attribute repair for two vehicular 
    telematic gateways.}
    \label{tab:ttcontrol}
    \begin{tabular}{lrrr}
    \toprule
        & \multicolumn{2}{c}{with flow} & without flow \\
        \cmidrule(r){2-3}
        \cmidrule(l){4-4}
         &  simple & full & full \\
\midrule
verdict     & SAT & SAT & SAT  \\
    total \# formulas & \change{95} & \change{95} & \change{82}  \\
    \# repairable formulas & \change{15} & \change{19} & \change{18}  \\
    \# unrepairable formulas & \change{25} & \change{23} & \change{21} \\
    \# formulas w/t threat & \change{55} & \change{53} & \change{43} \\
    total cost & \change{29} & \change{57} & \change{57} \\
    time (s) & \change{127} & \change{497} & \change{118} \\
    \bottomrule
    \end{tabular}
    \vspace{-15pt}
\end{table}

\change{
We illustrate one threat that the tool found repaired in the full model.

    \noindent \emph{Spoofing sensors by external effects:} This threat is present because an interface (in our case CAN) is connected to an ECU (in our case the secondary CPU) and they both hold an asset (in our case the Communication Interface). This threat represents the possibility that the assets could be attacked to send incorrect data to vehicle sensors (i.e., radar signals). That could lead to giving incorrect decisions based on the tampered input signal, affect the safe operation of the vehicle, or impact on usual vehicle functionalities. Our implementation suggests a repair for this threat by implementing an input validation on the Secondary CPU. The input validation enable the CPU to detect false (e.g. impossible) data.
}

\vspace{-10pt}
\vspace{-10pt}
\section{Related work}
\subsubsection{Threat modeling and analysis}

has received increasing interest in the recent years, both in academia and industry. A plethora of commercial and open-source threat modeling \change{tools} have been developed, including THREATGET~\cite{DBLP:conf/safecomp/SadanySK19,DBLP:journals/corr/abs-2107-09986}, 
Microsoft Threat Modeling Tool~\cite{mcree2014microsoft}, ThreatModeler~\cite{threatmodeler}, 
OWASP Threat Dragon~\cite{threatdragon} and pytm~\cite{pytm}, 
Foreseeti~\cite{foreseeti}, 
Security Compass SD Elements~\cite{sdelements} and Tutamen Threat Model Automator~\cite{tutamen}.
\change{
These tools can be divided into three categories: (1) manual approaches based on excel sheets or questionnaires~\cite{tutamen}, (2) graphical modelling approaches \emph{without} an underlying formal model~\cite{mcree2014microsoft,threatdragon,threatmodeler,sdelements}, and (3) model-based system engineering tools \emph{with} an underlying formal model~\cite{foreseeti,pytm,DBLP:conf/safecomp/SadanySK19,DBLP:journals/corr/abs-2107-09986}. The first class of tools does not admit automated threat analysis and any threat prevention measure must be manually identified and selected. 
Several tools from the second and third class~\cite{mcree2014microsoft,foreseeti,DBLP:conf/safecomp/SadanySK19,DBLP:journals/corr/abs-2107-09986} provide some limited form of non-automated threat mitigation by proposing a set of  hard-coded measures that are associated to individual threats or assets. 
However, these tools do not take into account potential threat inter-dependencies nor mitigation costs and are not able to compute a global and consistent set of threat prevention measures.
We note that our automated threat prevention approach is not applicable to the first class of tools. It could be integrated to the tools from the second and the third class. The integration to the second class of tools would likely be hard due to the lack of the underlying formal model on which our algorithm relies.

}
\subsubsection{Optimization Modulo Theories (OMT)}
 combine SMT solvers  with optimization procedures~\cite{DBLP:conf/tacas/BjornerPF15,DBLP:conf/sat/CimattiGSS13,DBLP:conf/cav/DilligDMA12,DBLP:conf/sat/NieuwenhuisO06,DBLP:journals/jar/SebastianiT20}. to find solutions optimizing some objective function 
%
Parameter synthesis using SMT solvers does not require optimization objectives in general. 
Bloem et al. \cite{DBLP:conf/cpsweek/RienerKFB16} synthesized parameter values ensuring safe behavior of cyber-physical systems through solving an $\exists\forall$ SMT formula.
In this work, users specify safe states in terms of state and parameter values; then, the synthesizer attempts to compute correct parameter values conforming to an invariant template such that for all possible inputs all reachable states are safe.


\section{Discussion and Future Work}
\label{sec:conclusion}


\change{
We presented a framework that 
enables automated threat prevention by repairing 
security-related system attributes.
Although widely applicable, attribute-value 
repair is not sufficient to cover all interesting 
security preventive procedures. For example, 
protecting safety-critical components connected to 
a Controller Area Network (CAN) bus in a vehicle 
cannot be done just by encrypting sensitive messages. 
In fact, encryption is not part of the CAN protocol. 
The preventive measure would require separating 
trusted (safety-critical) part of the system 
from the untrusted one (entertainment system, etc.) 
and putting a firewall in between -- an action that 
is beyond the attribute-value repair.
Despite few similar examples, the attribute repair remains 
a suitable repair strategy for the majority of threats present 
in common architectures. Intuitively, this is the case because 
the attributes document the counter-measures taken against common classes of threats, e.g. authentication as a counter-measure against escalation of privilege.
We plan to study the problem of threat prevention beyond attribute 
repair. The more general {\em model repair} problem can 
address the limitations illustrated by the CAN 
example from the introduction by enabling addition and removal of elements 
in a system architecture. Unrestricted alteration 
of system architecture models would lead to 
trivial and uninteresting repairs -- it suffices 
to disconnect all elements in the model from each other 
to disable the vast majority of threats. It 
follows that model repair requires a restriction 
of repair operations and even just identifying 
a set of useful 
repair operations is a challenging task.  
Although exciting, the more general model repair is 
a separate research problem that differs in several 
key aspects from the attribute-value repair and 
we plan to tackle it as future~work.
}



\subsubsection{Acknowledgments}
This research has received funding from the  European Union’s  Horizon  2020  research  and  innovation  programme  under  grant  agreement No 956123 and from the program ``ICT of the Future'' of the Austrian Research Promotion Agency (FFG) and the Austrian Ministry for Transport, Innovation and Technology under grant agreement No. 867558 (project TRUSTED).

\bibliographystyle{abbrv}
\bibliography{main}

\newpage
\appendix

\section{System Model}
\label{appendix:model}

This section defines a formal  model that we use to analyze a system against a database of threats. A system model provides an architectural view of the system under investigation, representing relevant components, their properties, as well as relations and dependencies between them.
\change{Complete, formal definitions of the system and the threat model are given in \cite{DBLP:journals/corr/abs-2107-09986}.}


The system model is designed from a set of available entities, 
such as {\em elements}, {\em connectors}, {\em security assets} and {\em security boundaries}. These entities are typed, can have attributes (properties) assigned to values drawn from some domain. 
The entity types, attributes and domains are not arbitrary -- they are selected from a predefined ``arena'' that we call a {\em meta model}. 

Before providing a formal definition of a meta model, we introduce useful notation. We use the term {\em item} to refer to a generic model entity when distinction is unnecessary.
In particular, we denote by 
$\itemuniverse = \elementuniverse \cup \connectoruniverse \cup \assetuniverse 
\cup \boundaryuniverse$ the universe of {\em items}, where $\elementuniverse$, 
$\connectoruniverse$, 
$\assetuniverse$ and 
$\boundaryuniverse$ are disjoint sets of all possible {\em elements}, {\em connectors}, {\em assets} and {\em boundaries}, respectively.  

In our framework, every item has a {\em type}. We denote by $\elementtypes$, $\connectortypes$, $\assettypes$ and $\boundarytypes$ the sets of valid types for elements, connectors, assets and boundaries, resp. We also use the notation 
$\itemtypes = \elementtypes \cup \connectortypes \cup \assettypes \cup \boundarytypes$ 
to group the types for generic items.

All items have {\em attributes} and attributes are assigned {\em values}. Let $\attributeuniverse$ denote the set of all attributes, and $\attributevalueuniverse$ the set of all attribute values. The mapping $\attributedomain~:~\attributeuniverse \to 2^{\attributevalueuniverse}$ defines the {\em domain} $\attributedomain(\attribute) \subseteq  \attributevalueuniverse$ of the attribute $\attribute \in \attributeuniverse$.


\begin{definition}[Meta model]
\label{def:meta}
A {\em meta model} $\metamodel$ is a tuple $\metamodel = (\itemtypes, \attributeuniverse, \attributevalueuniverse, \attributedomain, \itemattribute)$, where
\begin{itemize}
    \item $\itemtypes$, $\attributeuniverse$ and $\attributevalueuniverse$ are finite sets of \emph{item types}, \emph{attributes} and \emph{attribute values},
    \item $\attributedomain$ is a \change{function mapping attributes to sets of values}, and
    \item $\itemattribute~:~\itemtypes \to 2^{\attributeuniverse}$ is an {\em attribute labelling} that maps every item type $i\in \itemtypes$ to a finite set $\itemattribute(i) \subseteq \attributeuniverse$ of attributes.
\end{itemize}
\end{definition}

We are now ready to define a {\em system model} $\threatmodel$, which is instantiated by entities selected from its associated meta model $\metamodel$. 
Definition~\ref{def:system} formalizes a system model.

\begin{definition}[System Model]
\label{def:system}
A {\em system model} $\threatmodel$ is a tuple $\threatmodel=(\metamodel, \items, \type, \src, \tgt,  \attributevalueassignment, \attributeweight, \boundaryrelation, \assetrelation)$, where:
\begin{itemize}
    \item $\metamodel$ is a meta model,
    
    \item $\items \subseteq \itemuniverse$ is a finite set of {\em items}, partitioned into a disjoint union of 
    {\em elements} $\elements \subseteq \elementuniverse$, 
    {\em connectors} $\connectors \subseteq \connectoruniverse$,
    {\em assets} $\assets \subseteq \assetuniverse$, and
    {\em boundaries} $\boundaries \subseteq \boundaryuniverse$ ,
    
    \item $\type~:~I \to \itemtypes$ is the {\em type labelling} function that maps items $i \in I$ to their types $\type(i) \in \itemtypes$, such that
    (1) if $i\in \elements$, then $\type(i) \in \elementtypes$,
    (2) if $i \in \connectors$, then 
    $\type(i) \in \connectortypes$, 
    (3) if $i \in \assets$ then $\type(i) \in \assettypes$, 
    and (4) if $i \in \boundaries$ then $\type(i) \in \boundarytypes$,
    
    \item $\src~:~C \to E$ and $\tgt~:~C \to E$ denote functions that map a connector to its {\em source} and {\em target} elements, respectively.
    
    
    \item $\attributevalueassignment~:~I \times \attributeuniverse \rightharpoonup \attributevalueuniverse$ is a partial {\em attribute valuation} function, where (1) $\attributevalueassignment(i,\attribute)$ is defined if $\attribute \in \itemattribute(\type(i))$, and (2) if $\attributevalueassignment(i,\attribute)$ is defined, then $\attributevalueassignment(i,\attribute) \in \attributedomain(\attribute)$.
    
    \item\change
    {
        $\attributeweight~:~I \times \attributeuniverse \times \attributedomain \times \attributedomain \rightharpoonup \mathbb{R}_{\geq 0}$ is a partial {\em attribute cost} function such that $\attributeweight(i,\attribute,x,x')$ is defined only if $\attributevalueassignment(i,\attribute)$ is defined and $x,x'\in \domainofattribute(\attribute)$, which denotes the cost of changing the attribute $\attribute$ from value $x$ to $x'$ for item $i$.%
    }
    
    \item $\boundaryrelation \subseteq B \times (B \cup E)$ is a binary {\em boundary containment} relation that encodes a tree in which the leaves are elements of $E$ and the internal nodes are elements of~$B$. We use $\boundaryrelation^{*}$ to denote the transitive closure of the boundary containment relation.
    
    
    \item $\assetrelation~:~(E \cup C) \times A$ is the {\em asset relation} between~assets~on one side, and elements and connectors on the other side. Each element/connector can hold multiple assets. Similarly, each asset can be associated to multiple elements and connectors.

\end{itemize}
\end{definition}

\begin{definition}[Path]\label{def:path}
\change{
Given a system model $\threatmodel$, a path in $\threatmodel$ is an alternating sequence of elements and connectors in the form of $\pi = e_{1}, c_{1}, e_{2}, c_{2} \cdots, c_{n-1},e_{n}$, such that for all $1 \leq i \leq n$, $e_{i} \in E$, for all $1 \leq i  < n$, $c_{i} \in C$, $\src(c_{i}) = e_i$, and $\tgt(c_{i}) = e_{i+1}$ and for all $1 \leq i < j \leq n$, $e_{i} \neq e_{j}$. \footnote{We only consider acyclic paths. This is sufficient to express all interesting threats.}}
\end{definition}

We use the notation $\pathelms(\pi)$ and $\pathcons(\pi)$ to 
define the sets of all elements and of all connectors appearing in a path, respectively. The starting and the ending element in the path $\pi$ are denoted by $\elementstart(\pi) = \src(c_{1})$ and $\elementend(\pi) = \tgt(c_{n-1})$, 
respectively. 
We denote by $P(\threatmodel)$ the set of all paths in $\threatmodel$.

\begin{example}
We illustrate the system model on the generic IoT example from Section~\ref{sec:intro}. 
The model consists of $4$ elements (Cloud, Fog Node, IoT Field Gateway, Intelligent Monitoring Device) and $6$ connectors. Table~\ref{tab:type-label:appendix}
associates elements to their respective types, drawn from the meta model. There is only one type, Communication Flow, associated to all connectors.

\begin{table}
    \setlength{\aboverulesep}{1pt}
    \setlength{\belowrulesep}{1pt}
    \caption{Type Labelling}
    \label{tab:type-label:appendix}
    \centering
    \begin{tabular}{ll} 
    \toprule
    \multicolumn{1}{c}{\bf element} & \multicolumn{1}{c}{\bf type} \\ 
    \midrule
    Cloud & Cloud  \\ 
    Fog Node & Fog \\
    IoT Field Gateway & Gateway \\ 
    Intelligent Monitoring & Device \\
    \bottomrule
\end{tabular}
\end{table}
\end{example}

Every element and connector has multiple associated attributes. For instance attributes Authentication, Authorization and Tamper Protection are three attributes of the IoT Field Gateway element. Cloud elements have additional attributes, such as the Data Logging and Data Encryption. 
Data Encryption attribute can for instance take values from the domain $\{ \text{Yes}, \text{No}, \text{Undefined} \}$. 
\section{Threat Logic}
\label{appendix:logic}

We introduce {\em threat logic} for specifying potential threats. 
While the authors of THREATGET define their own syntax and semantics to express threats \cite{DBLP:journals/corr/abs-2107-09986} we use standard predicate logic in  this paper. 
The use of the predicate logic facilitates the 
encoding of the forthcoming algorithms into 
SMT formulas.
Our implementation contains an automated 
translation from THREATGET syntax to threat logic.

Let $\itemvars$ 
be a finite set of \emph{item} variables and 
$\pathvars$ a finite set \emph{path} variables. 
The syntax of threat logic is defined as follows:
$$
\begin{array}{lcl}
\varphi &  := & R(\itemvars \cup \pathvars)~|~\neg \varphi~|~\varphi_1 \vee \varphi_2~|~\exists \pathvar. \varphi~|~\exists \itemvar. \varphi \\
\end{array}
$$
\noindent where $\itemvar \in \itemvars$, 
$\pathvar \in \pathvars$, and 
$R(\itemvars \cup \pathvars)$ is a predicate of the 
form:
$$
\begin{array}{lclr}
R(\itemvars \cup \pathvars) & := & \ttype(x) = t  & (1)\\
                            &   & |~x_{\text{ec}} \isin p & (2) \\
                            &   & |~\hascon(x_e, x_c) & (3) \\
                            &   &  |~\source(x_c) = x_e & (4) \\
                            &   & |~\target(x_c) = x_e & (5) \\
                            &   &  |~\source(p) = x_e & (6) \\
                            &   & |~\target(p) = x_e & (7) \\
                            &   & |~\crosses(x_c, x_b) & (8) \\
                            &   & |~\contains(x_\mathit{eb}, x_b) & (9) \\
                            &   & |~\holds(x_{\mathit{ec}}, x_a) & (10) \\
                            &   & |~\attval(x, \emph{att}) = y & (11) \\
\end{array}
$$
where $x \in \items$, $x_e\in\elements$, $x_c\in\connectors$, $x_{\mathit{ec}}\in\elements\cup\connectors$, $x_b\in\boundaries$, $x_\mathit{eb}\in\elements\cup\boundaries$, $x_a\in\assets$, $\pathvar \in \pathvars$, $t \in \itemtypes$, $\emph{att} \in \attributeuniverse$, and $v \in \domainofattribute(\emph{att})$.


The predicate (1) holds if the type of $x$ is $t$, (2) holds if the element or the connector 
$x_{\text{ec}}$ is in the path $p$, (3) holds if the element 
$x_e$ is either the source or the target of the connector 
$x_c$, (4) holds if the 
source of the connector $x_c$ is the element $x_e$, 
(5) holds if the target of the connector $x_c$ is the element $x_e$, 
(6) holds if the 
source of the path $\pathvar$ is the element $x_e$, 
(7) holds if the target of the path $p$ is the element $x_e$,
(8) holds if the connector $x_c$ crosses the boundary $x_b$, (9) holds if the element or boundary $x_\mathit{eb}$ is contained in the boundary $x_b$, 
(10) holds if the element or the connector $x_{\mathit{ec}}$ 
holds the asset $x_a$, and (11) holds if the valuation of the attribute {\em att} 
associated to the item $x$ is equal to $y$.

\begin{example}
Consider a requirement that states that there exists a  
path in the model such that all the elements in that path fulfil a certain conditions, e.g. all elements are of type Cloud. The following threat logic formula expresses the above requirement:
$$
\exists p.\forall e.(e \isin p \implies \ttype(e)= \text{Cloud}).
$$
\end{example}



\change{We define an \textit{assignment} 
$\projection_{\threatmodel}$ as a partial function 
that assigns item and path variables to items and paths from 
$\threatmodel$.}
We denote by $\projection_{\threatmodel}[\itemvar \mapsto i]$
the item assignment in which $\itemvar$ is mapped to $i$ and otherwise identical to \change{$\projection_{\threatmodel}$}.
Similarly, we denote 
by $\projection_{\threatmodel}[\pathvar \mapsto \pi]$
the path assignment in which $\pathvar$ is mapped to $\pi$ and otherwise identical to \change{$\projection_{\threatmodel}$}.
The semantics of threat logic are defined inductively in Table~\ref{tab:sem:appendix}.

\begin{table*}
\centering
\caption{Threat logic semantics.}
\label{tab:sem:appendix}
$\hfill
\begin{array}{lcl}

\assignment_{\threatmodel} \models \ttype(\itemvar) = t  & \leftrightarrow &  \type(\assignment_{\threatmodel} (\itemvar)) = t \\

\change{\assignment_{\threatmodel} \models \itemvar_{\text{ec}} 
\isin \pathvar} & \leftrightarrow & 
\change{\assignment_{\threatmodel} (\itemvar_{\text{ec}}) \in (\pathelms(\assignment_{\threatmodel}(\pathvar)) \cup 
\pathcons(\assignment_{\threatmodel}(\pathvar)))}\\

\assignment_{\threatmodel} \models \hascon(x_e, x_c) & \leftrightarrow & \src(\assignment_{\threatmodel}(x_c)) = 
\assignment_{\threatmodel}(x_e) \text{ or } 
\tgt(\assignment_{\threatmodel}(x_c)) = \assignment_{\threatmodel}(x_e) \\

\assignment_{\threatmodel} \models \source(x_c) = x_e & \leftrightarrow & 
\src(\assignment_{\threatmodel}(x_c)) = \assignment_{\threatmodel}(x_e) \\

\assignment_{\threatmodel} \models \target(x_c) = x_e & \leftrightarrow & \tgt(\assignment_{\threatmodel}(x_c)) = 
\assignment_{\threatmodel}(x_e)\\

\change{ \assignment_{\threatmodel} \models \source(p) = x_e} & \leftrightarrow & \change{
\elementstart(\assignment_{\threatmodel}(p)) = \assignment_{\threatmodel}(x_e) } \\

\change{\assignment_{\threatmodel} \models \target(p)} = x_e & \leftrightarrow & \change{ \elementend(\assignment_{\threatmodel}(p)) = 
\assignment_{\threatmodel}(x_e) }\\

\assignment_{\threatmodel} \models \crosses(x_c, x_b) & \leftrightarrow & (\assignment_{\threatmodel}(x_b), \src(\assignment_{\threatmodel}(x_c))) \not \in \boundaryrelation^{*}  \Leftrightarrow (\assignment_{\threatmodel}(x_b), \tgt(\assignment_{\threatmodel}(x_c))\change{)} \in \boundaryrelation^{*}\\

\assignment_{\threatmodel} \models \contains(x_\mathit{eb}, x_b) & \leftrightarrow & (\assignment_{\threatmodel}(x_b), \assignment_{\threatmodel}(x_{eb})) \in \boundaryrelation^{*} 
\\


\assignment_{\threatmodel} \models \holds(x_{\text{ec}}, x_a) & \leftrightarrow & (\assignment_{\threatmodel}(x_{\mathit{ec}}), \assignment_{\threatmodel}(x_a)) \in \assetrelation\\ 

\assignment_{\threatmodel} \models \attval(x, \emph{att}) = y & \leftrightarrow & 
\attributevalueassignment(\assignment_{\threatmodel}(x), \emph{att}) = y \\

\assignment_{\threatmodel} \models \neg \varphi & \leftrightarrow & \assignment_{\threatmodel} \not \models \varphi \\

\assignment_{\threatmodel} \models \varphi_{1} \vee \varphi_{2} & \leftrightarrow & \assignment_{\threatmodel} \models \varphi_{1} \text{ or } \assignment_{\threatmodel} \models \varphi_{2} \\

\assignment_{\threatmodel} \models \exists \itemvar. \varphi & \leftrightarrow & \exists i \in \items.
\assignment_{\threatmodel}[\itemvar \rightarrow i] \models \varphi \\

\change{\assignment_{\threatmodel} \models \exists \pathvar. \varphi} & \leftrightarrow & \change{\exists \pi \in {P}(\threatmodel).
\assignment_{\threatmodel}[\pathvar \rightarrow \pi] \models \varphi} \\

\end{array}
$\hfill\mbox{}
\end{table*}

We say that a threat logic formula is {\em closed} when all occurrences of item variables 
are in the scope of a quantifier. Any closed threat logic formula is a valid threat 
specification. Given a system model $\threatmodel$ and a closed threat logic 
formula $\varphi$, we say that $\threatmodel$ {\em witnesses} the threat $\varphi$, 
denoted by $\threatmodel \models \varphi$ iff $\assignment_{\threatmodel} \models \varphi$, 
where $\assignment_{\threatmodel}$ is an empty assignment.

\subsection{From Threat Logic To First Order Logic}
\label{appendix:fol}

\change{We observe that we interpret threat logic formulas over system models with a finite number of elements and connectors, and hence we can eliminate path quantifiers by enumerating the elements and the connectors in the path.}

\changenew{
$F_{\threatmodel}$ is the translation of the model ${\threatmodel}$, where the  elements, connectors, boundaries, and assets are encoded as enum sorts. Further, $F_{\threatmodel}$ consists of the functions $\type$, $\src$, $\tgt$ as well as the relations $\assetrelation$ and $\boundaryrelation^{*}$. The function $\attributevalueassignment$ is not part of $F_{\threatmodel}$, because the attribute values are not hard assertions. Instead we add the following constraints to $F_{\threatmodel}$ to ensure attributes can have only valid values:

\begin{align*}
&\forall \itemvar\in\items, \attribute\in\attributeuniverse.\ (\type(\itemvar)\notin\itemattribute(\attribute)\Rightarrow\attributevalueassignment(x, \attribute) = \bot)\ \wedge \\
& \forall \itemvar\in\items, \attribute\in\attributeuniverse.\ (\type(\itemvar)\in\itemattribute(\attribute)\Rightarrow\ \bigvee_{y\in\attributedomain(\attribute)} \attributevalueassignment(x, \attribute)=y)
\end{align*}

The set of soft assertions $\Psi = \{F_{1}, \ldots, F_{m}\}$ is defined as 
\[\{F_{i,\attribute,x}\}_{i\in\items, \type(\itemvar)\in\itemattribute(\attribute),x \in \domainofattribute(\attribute),x \neq \attributevalueassignment_\iota(i,\attribute)},\]
where $F_{i,\attribute,x}=\bigvee_{x' \in \domainofattribute(\attribute), x\neq x'} \attributevalueassignment(i,\attribute)=x'$. 

We use $\attributevalueassignment_\iota$ to refer to the original attribute value assignment in $\threatmodel$ and $\attributevalueassignment$ for the value assignment the SMT solver discovers.
Each soft assertion is a disjunction over all possible attribute values with the exception of the possible new value. This disjunction is violated iff the attribute is indeed assigned this new value. Note that the current value of the attribute is in each disjunction and therefore keeping the current value of an attribute has cost 0.
The cost of $F_{i,\attribute,x}$ is \[\mathit{cost}(F_{i,\attribute,x})=\attributeweight(i,\attribute, \attributevalueassignment_\iota(i,\attribute), x).\]
Finally, we denote by $\hat{F}_{\threatmodel} = F_{\threatmodel} \wedge F_{1} \wedge \ldots 
\wedge F_{m}$ the full translation of the model $\threatmodel$ that includes the 
soft assertions. 
}

\change{

\begin{definition}
Consider a system model $\threatmodel$ with $n$ elements. We define the inductive translation operator $T$, \changenew{that translates threat logic formulas to first-order formulas with their known semantics. The result of $T$ can be fed directly into an SMT solver.}
Given a threat logic formula $\varphi$ defined over 
$\itemvars$ and $\pathvars$, we define a mapping 
$\sigma$ that associates to 
every $p \in P$ a set $\{x^{1}_{p}, \ldots, x^{n-1}_{p}, k_{p}\}$ of fresh and unique $x^{i}_{p}$ item (connector) variables not in $\itemvars$ and a fresh and unique $k_{p}$ integer variable. \changenew{We use $k_p$ to denote the length of the actual path discovered by the SMT solver and $n$ for the maximum possible length of a path, where $n$ is the number of elements in $\threatmodel$.}

We define
$$
\begin{array}{lcl}
T(\source(p) = x_{e}) & = & x_{e} = \src(x_{p}^{1}) \\
T(\target(p) = x_{e}) & = &  \bigvee_{i=1}^{n-1}k_p=i\wedge x_{e} =\tgt(x_{p}^{i})  \\
T(x_{\mathit{ec}} \isin p) & = & 
\bigvee_{i=1}^{n-1} (i \leq k_p \wedge (x_{\mathit{ec}} = x^{i}_{p} \vee  \\
&& x_{\mathit{ec}} = \src(x^{i}_{p}) \vee x_{\mathit{ec}} = \tgt(x^{i}_{p}))) \\
T(R'(X)) & = & R'(X) \\
T(\neg \varphi) & = & \neg T(\varphi) \\
T(\varphi_{1} \vee \varphi_{2}) & = & T(\varphi_{1}) \vee T(\varphi_{2}) \\
T(\exists x. \varphi) & = & \exists x. T(\varphi) \\
T(\exists p. \varphi) & =&  \exists x_{p}^{1}, \ldots, x_{p}^{n-1}\in\connectors, k_{p} \in \{1,\ldots,n{-}1\}. \\
& & \Big(\bigwedge_{i=1}^{n-1}\Big( i \leq k_{p} \Rightarrow \\
&(3)& (\bigwedge_{j=1}^{i} \tgt(x_{p}^{i}) \neq \src(x_{p}^{j}))\wedge \\
&(4) & (i = 1\vee\tgt(x_{p}^{i-1}) = \src(x_{p}^{i}))\Big)\Big) 
 \wedge T(\varphi) \\
\end{array}
$$
\noindent where $R'(X)$ are the other predicates defined over item variables only. 
\end{definition}
}


\changenew{
The translation of $T(\exists p. \varphi)$ ensures that $\{x^i_p\}_i$ and $k_p$ 
encode a path of size $k_p$ (see (4)) that is acyclic (i.e. no element 
in the encoded path is repeated, see (3)).
}

\change{
\begin{restatable}{proposition}{proptransnew}
\label{prop:trans:appendix}
Let $\threatmodel$ be a system model, $\assignment_{\threatmodel}$ an assignment function and $\varphi$ a closed threat logic formula. 
We have that
$$
\assignment_{\threatmodel} \models \varphi 
\text{ iff } \hat{F}_\threatmodel \wedge T(\varphi) \text{ is satisfiable} \\
$$
\end{restatable}
}

\begin{example}
\label{ex:rules:appendix}
We formalize in Table~\ref{tab:threats:appendix} the two threats described informally in Section~\ref{sec:motivating}. 

\begin{table}[H]
    \centering
    \caption{Two threats from motivating example  in threat logic.}
    \label{tab:threats:appendix}
    \begin{tabular}{ll}
    \toprule
    Threat 1     &  $\exists e. (\ttype(e) = \text{Cloud} \wedge \attval(e, \text{Data Logging}) = \text{Yes})$ \\
                &   $\wedge \attval(e, \text{Data Encryption}) \neq \text{Yes})$   \\
    
    Threat 2     &  \change{$\exists p, x^{1}_{e}, x^{2}_{e}. 
    \source(p) = x^{1}_{e} \wedge \target(p) = x^{2}_{e} \wedge$} \\
    &
    \change{$\ttype(x^{2}_{e}) = \text{Cloud} \; \wedge \ttype(x^{1}_{e}) = \text{Device} $} \\
                & \change{$\wedge \attval(x^{2}_{e}, \text{Data Logging}) \neq \text{Yes})$} \\
    \bottomrule
    \end{tabular}
\end{table}
\end{example}

\subsection{Proofs}
\label{app:proofs}

\changenew{
For the following proofs we recall our definition of $T(\exists p. \varphi)$.
\begin{align}
T(\exists p. \varphi)  =&  \exists x_{p}^{1}, \ldots, x_{p}^{n-1}\in \connectors, k_{p} \in \{1,\ldots,n{-}1\}. \label{line:vars} \\
& \Big(\bigwedge_{i=1}^{n-1} \Big(i \leq k_{p} \Rightarrow \label{line:bound} \\
& (\bigwedge_{j=1}^{i} \tgt(x_{p}^{i}) \neq \src(x_{p}^{j}))\wedge \label{line:unique} \\
& (i = 1\vee\tgt(x_{p}^{i-1}) = \src(x_{p}^{i}))\Big)\Big) \wedge T(\varphi) \label{line:connected}
\end{align}
}

\changenew{

\begin{proof}
The proof works by structural induction over the threat logic formula. 
The base cases are if $\varphi$ equals to:
\begin{itemize}
    \item $R'(X)$: since $T(R'(X)) = R'(X)$ and this directly refers to the semantics, the property is fulfilled
    \item $\source(p) = x_{e}$: since $T(\source(p) = x_e) = (\src(x_1^p)=x_e)$, the property is fulfilled
    \item $\target(p) = x_{e}$: The target of a path is the target of the last connector in the path. For a known length of the path $k_p$ the disjunction collapses to $x_{e} =\tgt(x_{p}^{k_p})$, because $k_p=i$ is false for all other $i$. This corresponds to the semantics of the path.
    \item $x_{\text{ec}} \isin p$: Using the same argument as above, all $x_p^i$ past the length of the path $k_p$ are ignored.
\end{itemize}

The induction cases are $\neg \varphi$, $\varphi_{1} \vee \varphi_{2}$, $\exists x. \varphi$ and $\exists p. \varphi$. All but the last case are trivial.

To prove the correctness of the translation of $\exists p.\varphi$ we split the problem into two directions.
In the first direction we proof that if there exists a path $p$ in our model our first order logic encoding will find it. Suppose there exists a path $\pi \in {P}(\threatmodel) = e_{1}, c_{1}, e_{2}, c_{2} \cdots, c_{l-1},e_{l}$, such that 
$\assignment_{\threatmodel}[\pathvar\rightarrow\pi] \models \varphi$.
We show that then $\hat{F}_\threatmodel \wedge T(\exists p. \varphi) \text{ is satisfiable}$ by giving values for the variables $x_{p}^{1}, \ldots, x_{p}^{n-1}, k_{p}$:

\[
\begin{array}{lcl}
     k_p & = & l-1  \\
     x_{p}^{1} & = & c_1 \\
     \vdots \\
     x_{p}^{l-1} & = & c_{l-1} \\
     x_{p}^{l} & = & c_1 \\
     \vdots \\
     x_{p}^{n-1} & = & c_1
\end{array}
\]

The condition $i \leq k_{p}$ ensures that all $x_{p}^{i}$ for $i > l-1$ are ignored in the following checks.
For $1\leq i < j \leq l-1$ condition (\ref{line:unique}) is clearly fulfilled as $\pi$ is an acyclic path. Condition (\ref{line:connected}) follows because $c_{1}, \ldots, c_{l-1}$ form a path. $T(\varphi)$ follows by induction.

For the second direction we assume an assignment to $x_{p}^{1}, \ldots, x_{p}^{n-1}, k_{p}$ fulfilling conditions (\ref{line:vars}-\ref{line:connected}) and we need to show that there exists a path $\pi \in {P}(\threatmodel)$, such that 
$\assignment_{\threatmodel}[\pathvar\rightarrow\pi] \models \varphi$.
We construct the path $\pi = \src(x_{p}^{1}) x_{p}^{1} \tgt(x_{p}^{1}) x_{p}^{2} \tgt(x_{p}^{2})\ldots x_{p}^{k_p} \tgt(x_{p}^{k_p})$. This is a valid path because $\src(x_{p}^{i})=\tgt(x_{p}^{i-1})$ for all $1<i\leq k_p$ due to condition (\ref{line:connected}). It is also an acyclic path due to condition (\ref{line:unique}). $\varphi$ follows by induction.
\end{proof}

}

\thmmain*

\change{
\begin{proof}
Given a system model $\threatmodel$, its encoding $F_\threatmodel \land \bigwedge_{F \in \Psi} F$ is satisfiable by definition. Suppose a set of threat rules $\{ \varphi_1, \ldots, \varphi_n\}$, we have three cases:

\begin{enumerate}
    \item If $\threatmodel \models \bigwedge_{i=1}^{n} \neg \varphi_{i}$ then 
    $F_\threatmodel \land \bigwedge_{F \in \Psi} F \land \bigwedge_{i=1}^{n} \neg T(\varphi_{i})$ is satisfiable.

    \item If $\threatmodel \not \models \bigwedge_{i=1}^{n} \neg \varphi_{i}$ and $\solve(F_\threatmodel \land \bigwedge_{i=1}^{n} T(\neg \varphi_{i}))=\unsat$, then there exists $\varphi_i$ that cannot be repaired and reveals a structural threat regardless of the attributes.
    
    \item If $\threatmodel \not \models \bigwedge_{i=1}^{n} \neg \varphi_{i}$ and $\solve(F_\threatmodel \land \bigwedge_{i=1}^{n} T(\neg \varphi_{i}))=\sat$ then $\exists \Psi' \subseteq \Psi$ s.t. $\solve(F_\threatmodel \land \bigwedge_{i=1}^{n} T(\neg \varphi_i) \land \bigwedge_{F \in \Psi'} F)=\unsat$, and $\maxsolve(F_\threatmodel \land \bigwedge_{i=1}^{n} T(\neg \varphi_i) \land \bigwedge_{F \in \Psi'} F)=\sat$ with some cost $k$ greater than zero.
\end{enumerate}

Consider case (3). Suppose that the minimum attribute threat repair for $M$ (with valuation function $v$) and $\{\varphi_{1}, \ldots, \varphi_{n}\}$ has cost $k' < k$. By definition of minimum attribute threat repair, there exists a valuation function $v'$, such that $\threatmodel[v'\backslash v] \models \wedge_{i=1}^{n} \neg \varphi_i$ and $d(v,v') = k'$. Let $\Psi' = \{F_{i, \attribute, x}~|~v(i,\attribute) \neq v'(i, \attribute)\}$ be the set of soft assertions for the 
item-attribute pairs $(i,\attribute)$ for which the valuation changes from $x$ to $x'$ with $\mathit{cost}(F_{i, \attribute, x'}) = w(i,\attribute, x,x')$, i.e. $\Sigma_{F \in \Psi'} \mathit{cost}(F) = k'$. It follows that $\solve(F_\threatmodel \land \bigwedge_{i=1}^{n} T(\neg \varphi_i) \land \bigwedge_{F \in \Psi\backslash \Psi'} F)=\sat$, hence 
$\maxsolve(F_\threatmodel \land \bigwedge_{i=1}^{n} T(\neg \varphi_i) \land \bigwedge_{F \in \Psi'} F)=\sat$ with cost $k'$, which is a contradiction. It follows that MaxSAT provides the 
solutions to the minimum attribute repair problem.

\end{proof}
}

\section{Case Study: Key Fob}
\label{sec:key:appendix}

\begin{figure} 
  \centering
  \input{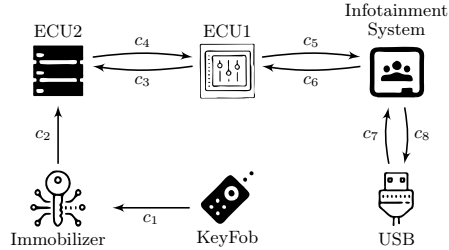} 
  
\caption{Example modeled in THREATGET}\label{fig:KeyFob}
\end{figure}

This case study consists of a simplified architectural model of a remote locking and unlocking mechanism in a car, depicted in Figure~\ref{fig:KeyFob}. This model illustrates the main capabilities of the THREATGET tool. It consists of a key fob that communicates in a wireless fashion with the car's lock/unlock system.

\noindent This system is connected to an ECU (Electronic Control Unit) (ECU2 in Figure~\ref{fig:KeyFob}) that implements the actual locking and unlocking functionality. After processing the signal, the user gets feedback (e.g., a flashing light). The ECU realizing the lock/unlock feature is connected (via a CAN bus) to an ECU that controls the Infotainment System. The Infotainment System interacts with the driver  via a USB port. Components (elements) and connectors in the model have  attributes associated to them, and these attributes have values. For instance, the connector between the Key Fob and the Lock/Unlock system has an Encryption attribute set to {\em false}, meaning that no encryption is used to communicate between these two elements. 
Threats are structural security weaknesses in the model that are modeled as relations between entities in the model  (elements, connectors, attributes and their values, etc.). A potential threat in the remote locking model could be the presence of unprotected (non-encrypted) wireless communication channels. 
The model has a total of 6 elements and 8 connectors, without any assets or security boundaries.

\begin{table}
    \centering
    \caption{Results of attribute repair applied to the Key Fob case study.}
    \label{tab:key}
    \begin{tabular}{lrrrr}
    \toprule
        & \multicolumn{2}{c}{All threats} & \multicolumn{2}{c}{Subset of threats}\\
       \cmidrule(r){2-3}
    \cmidrule(l){4-5}
         &  full & h-partial & full & h-partial \\
        \midrule
    verdict     & UNSAT & SAT & SAT & SAT \\
    total \# formulas & \change{165} & \change{165} & 21 & 21 \\
    \# repairable formulas & n/a & \change{25} & 4 & 4 \\
    \# unrepairable formulas & n/a & \change{7} & 0 & 0 \\
    \# formulas w/t threat & n/a & \change{133} & 17 & 17\\
    total cost & n/a & \change{33} & 9 & 11 \\
    time (s) & 3.65 & \change{103} & 9.58 & 25.65 \\
    \bottomrule
    \end{tabular}
\end{table}

\change{
The threat repair procedure results in 
an optimized configuration of the 
system architecture model with security 
attributes that prevent threats from the 
original model. The implementation of the 
recommended measures has an immediate 
positive impact on the system security. We 
illustrate one of the  
proposed prevention measures.

\noindent \emph{Manipulation of vehicular data:} this threat is present because a malicious user could 
    connect a Linux machine to the USB interface, 
    access the CAN bus via the Infotainment system, 
    manipulate command data and send them to the 
    ECUs through the CAN bus. The recommended 
    prevention 
    measure consists in implementing 
    authentication for connections to the USB interface, 
    which would result in ignoring 
    unauthenticated command messages. 

}

We summarize the evaluation of this case
in Table~\ref{tab:key}.
We first tried the vanilla procedure (Algorithm~\ref{alg:full}). 
The procedure gave an UNSAT verdict in 3.65s, despite the fact that all unrepairable rules (according to the has\_attr() syntactic check) were removed from the optimization. This result indicates that that there are a-priori repairable threat formulas that are either inconsistent with the system model or with another threat formula. We note that the UNSAT outcome is not very useful to the engineer because it does not give any (even partial) repair.
On the other hand, the heuristic method (Algorithm~\ref{alg:approximate}) successfully computed a partial repair of the system model, repairing \change{25} out \change{32} threats in \change{103s}.

In the next step, we manually selected 21 threat formulas 
that could be simultaneously repaired by the vanilla method. 
The vanilla procedure found 4 threats and repaired all of them 
in 9.58s with the cost on 9. The heuristic method also found 
4 threats, repaired all of them, but with the sub-optimal 
cost of 11 in 25.65s. We can see that the flexibility of 
the heuristic method comes at a price - both in terms of 
optimality of the solution and computation time.
\section{Case Study: Vehicular Telematic Gateway}
\label{sec:ttcontrol:appendix}

\label{appendix:vtg}
\begin{figure*} 
\newpage
\vfill
    \centering
    \rotatebox{90}{%
        \resizebox{0.975\textheight}{!}{\input{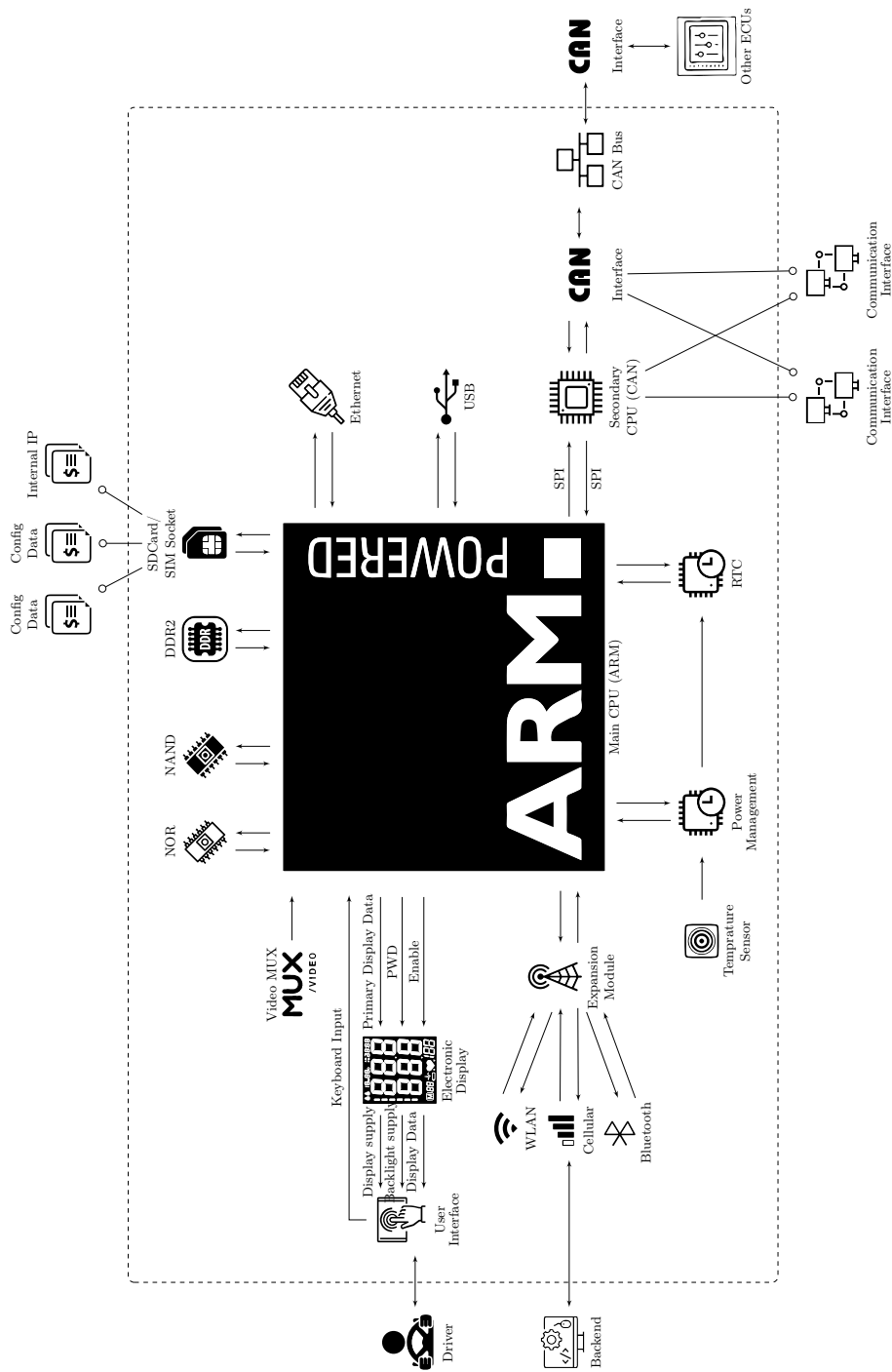}}%
    }
    \caption{Model of Vehicular Telematic Gateway}\label{fig:TTcontrol}
\newpage
\end{figure*}

\end{document}